# Chasing beams and muon colliders

G. Mambriani and R. Coïsson

Dipartimento di Fisica e Scienze della Terra , Università di Parma, 43100 Parma, Italy

giuseppe.mambriani@fis.unipr.it , roberto.coisson@fis.unipr.it

## Abstract

A possible alternative way of producing muons for $\mu^+ - \mu^-$ colliders (or other unstable particles) is proposed. It consists in colliding beams in a "chasing beam" configuration, i.e. collisions of two beams having the same direction but with different energies. This would produce muons with a good collimation, having already a high energy from the beginning, and then a longer life.

In the search for higher and higher energies, electron-positron circular colliders have a limit in the power dissipated by synchrotron radiation, which strongly increases with energy. One of the solutions that have been envisaged is to design linear colliders; the other one, as suggested long ago by Budker [1,2], is to use muons, which, due to their larger mass, emit much less synchrotron radiation, being otherwise - for the physics to be studied - just like electrons. For recent reviews of the feasibility studies of muon colliders, see [3-7]. There is a fundamental problem with muons, however, that their lifetime is limited. The method that has been proposed to produce muons to be accelerated is to send protons on a fixed target, and then accelerate and collimate pions which then decaying form the muon beams. Here we suggest a possible alternative for the production of the muon beams, which consists in colliding proton and antiproton (or proton) beams, both travelling in the same direction with different energies, a configuration which can be called "chasing beams" The advantage of this would be that the center of mass of the collision has a high velocity, so that the muons are created already with a high Lorentz factor, and then have a longer life (in terms of laboratory time). Another advantage is a relatively good collimation due to the Lorentz trasformation of angles from the center of mass to the laboratory frame. We just remind some simple formulas of special Relativity for the discussion of possible parameters.

If the slower particle has a Lorentz factor $\gamma_2 = (1 - \beta_2/c^2)^{-1/2}$, if the chasing particle must have a relative velocity $\beta_R$ with respect to the chased one, its Lorentz factor $\gamma_2$ is such that

$$\beta_2 = (\beta_1 + \beta_R)/(1 + \beta_1 \beta_R)$$

If the muon, with rest mass m, is subject to a force F and starts with initial value $\beta_0$, $\gamma_0$, its proper elapsed time $\tau_f$ when it reaches $\beta_f$ and $\gamma_f$ can be found by taking into account that $dt/d\tau = \gamma$ and $F/mc = dw/dt$, where $w = \beta\gamma = (\gamma^2 - 1)^{1/2}$, and then calling $\eta$ the dimensionless impulse $Ft/mc$, $\eta = w - w(0)$, then $F/mc*\tau = d\eta/\gamma$, or, if we start at $t = \tau = 0$,

$$\tau = \frac{F}{mc} \int_0^{\eta_f} (1 + (\eta + w_0)^2)^{-1/2} = \log((w_f + (1 + w_f^2))^{-1/2}/(w_0 + \gamma_0))$$

This gives the proper elapsed time of the muon as a function of its initial energy and the average accelerating force. For example, we can see that for an accelerating field of 1 MV/m, with an initial energy of 100 GeV, at an energy of 500 GeV there are still 77.4% of surviving muons.

Let us consider two possible schemes, one based on proton-antiproton collisions, and one on proton-proton.

In proton-antiproton collisions producing pions, if the relative dimensionless speed $\beta_R$ is sufficiently low, the annihilation cross-section is expected to be proportional to $1/\beta_R$, that is $\sigma = A/\beta_R$. As at 300 MeV/c $\beta_R = 0.943$ the annihilation cross-section is approximately $1.5 \times 10^{-25}$ cm$^2$, we have $A = 1.4 \times 10^{-25}$ cm$^2$. If the first beam (say, of protons) has an energy of 50 GeV, the chasing beam could have an energy of 50.73 GeV $\beta_R = 0.014$ or 94 KeV, and a cross-section of $10^{-23}$ cm$^2$

In the collision rest system, if pions are emitted isotropically, about 10% of the charged pions are within 0.65 rad from the forward direction. Once transformed to the labotratory frame, this means a half-opening angle of 6 mrad. If the p and p beams are focussed to 5 $\mu$m, and their emittance is of the order of 10 $\pi$ nm.rad, the angular spread in the forward cone is of the order of the original beam divergence, so the emittance of the pion beams is not substantially increased with respect to the proton and antiproton beams. The decay into $\mu^+$ and $\mu^-$ does not happen at a specific point, so one cannot use any focussing to reduce the the effect of the increased angular spread, however the added angular spread in this decay is of the order of $(m_\pi - m_\mu)c/$ 95 GeV/c<1 mrad.

The other method to produce collimated pions and then muons could be that of using two chasing beams of protons, at a sufficient energy in the center-of-mass frame for having an intense pion yield, but with one of the beams having a much greater energy in the lab frame, and the other relatively smaller. For instance, if we want 1 GeV in the CM frame, and if the high-energy beam is chosen to have 50 GeV, the other beam must have an energy of nearly 6 GeV

The decay length for pions having a $\gamma$ of about 300 is nearly 3 Km, so an adequate pion decay region must be considered, which could be done in a circular tunnel.

As an example, we can think of using the SPS ring at CERN as an injector to a suitably modified LHC.

Once the pions are decayed and the muons focussed and filtered to get the suitable phase space volume, they could be injected into the accelerator. The pulse of muons must be subdivided into a suitable number of bunches for the acceleration.

Let us imagine we can get for each pulse $10^{15}$ protons against $10^{14}$ antiprotons, and a crossing area of $10^{-7}$ cm$^2$ and that (at the above considered cross-section of $10^{-23}$ cm$^2$ all the antiprotons are made to annihilate. One can take that it is possible to inject $10^{11}$ muons of both signs into the SPS. If the SPS was used as a final muon collider in the range 300-400 GeV (for ex. to study top-quark physics and toponium properties). Muons with $\gamma = 2 \times 10^3$ have a mean life in the lab frame of about 5 ms, during which they can make nearly 200 turns in the SPS ring. With four intersection regions, where muons are focussed to $10^{-7}$ cm$^2$, if we assume a useful cross-section of $10^{-35}$ cm$^2$, $10^{11} + 10^{11}$ muons per pulse, and 10 pulses per second, one gets roughly $10^4$ events per day. Imagining to do the same estimate for muons accelerated by the SPS and injected into the LHC up to 5 TeV energy, assuming a cross-section of $10^{-37}$ cm$^2$, $2 \times 10^{10}$ muons per beam, two interaction regions (and a free path at 5 TeV of $3 \times 10^4$ Km), one gets of the order of one event per day.

Both these schemes (p-p and p-pbar) could also be used for other purposes.

For example we could use the reaction

p+p=3p+pbar with cross-section ~ $10^{-23}$ cm$^2$ at ~2 GeV above its CM threshold.

With a pulse of $10^{15}$ protons at 50 GeV chasing a pulse of $10^{16}$ protons at 70 MeV (both focussed to $10^{-7}$ cm$^2$), one has roughly 2 GeV in the CM frame and $10^{15}$ antiprotons per pulse in the appropriate phase space volume.

On the other hand, the proton-antiproton scheme, besides muons, one could obtain several types of beams with unprecedented high intensities: charged and neutral pions, charged and neutral kaons, and lambdas and antilambdas.

Although at present the estimates of numbers of protons and antiprotons available in accelerators are not yet reached, this scheme could be interesting to keep in mind for the future.